\documentclass[12pt,twoside]{reedthesis}

\usepackage{graphicx,latexsym} 
\usepackage{amssymb,amsthm,amsmath}
\usepackage{longtable,booktabs,setspace,pdfsync,verbatim,listings} 
\usepackage{url}
\usepackage{natbib}
  \usepackage[charter]{mathdesign} % other fonts are available like times, bookman, charter, palatino

\title{Relativistic Springs}
\author{Dylan P. Clark}
\date{May 2011}
\division{Mathematics and Natural Sciences}
\advisor{Nelia Mann}

\department{Physics}

\begin{document}

  \maketitle
  \frontmatter % this stuff will be roman-numbered
  \pagestyle{empty} % this removes page numbers from the frontmatter

% Acknowledgements (Acceptable American spelling) are optional
% So are Acknowledgments (proper English spelling)
\chapter*{Acknowledgements}
To Nelia,\vspace{.1in}\\ Without your patient support this would have been impossible.\vspace{.25in}\\
To Joel and MJ.\vspace{.1in}\\
To Mom, Dad, Miles, and Teddy.\vspace{.1in}\\
To the Fishbowl.\vspace{.1in}\\
To Max and Todd.\vspace{.1in}\\
And to A. Genova.\vspace{.1in}\vspace{.25in}\\

% The preface is optional
% To remove it, comment it out or delete it.
%    \chapter*{Preface}
%	This is an example of a thesis setup to use the reed thesis document class.

    \enlargethispage{36pt}
    \tableofcontents
% if you want a list of tables, optional
 %   \listoftables
% if you want a list of figures, also optional
    \listoffigures

% The abstract is not required if you're writing a creative thesis (but aren't they all?)
% If your abstract is longer than a page, there may be a formatting issue.
    \chapter*{Abstract}
	Here we develop a model for the relativistic spring. We examine the effects of revising the simple harmonic oscillator to include relativistic momentum and a delayed force law. These corrections alter two of the most significant features of the simple harmonic oscillator: energy conservation and a constant period independent of initial conditions. The relativistic momentum correction, while preserving energy conservation, does not have period independent of initial conditions. The delayed force law, while preserving period independence, does not conserve energy. Applying both corrections creates a solution with increasing amplitude and increasing period, a result that is very different from the traditional simple harmonic oscillator. 

  \mainmatter % here the regular arabic numbering starts
  \pagestyle{fancyplain} % turns page numbering back on

%The \introduction command is provided as a convenience.
%if you want special chapter formatting, you'll probably want to avoid using it altogether

	%%%%%%%%%%%%%%%%%%%%%%%%%%%%%% INTRO %%%%%%%%%%%%%%%%%%%%%%%%%%%
	%%%%%%%%%%%%%%%%%%%%%%%%%%%%%% INTRO %%%%%%%%%%%%%%%%%%%%%%%%%%%
	%%%%%%%%%%%%%%%%%%%%%%%%%%%%%% INTRO %%%%%%%%%%%%%%%%%%%%%%%%%%%
\newcommand{\ttdd}[1]{\frac{d}{dt} \left \{ #1 \right \}}
\newcommand{\amma}{\sqrt{1-\left ( \frac{\dot x}{c} \right) ^2}}
\newcommand{\ttd}[1]{\frac{d #1}{dt}}
\newcommand{\gfrac}[1]{\frac{ #1}{\amma}}
\newcommand{\bs}[1]{\boldsymbol{ #1 }}

%\doublespacing
\onehalfspacing
    \chapter*{Introductory Remarks}
         \addcontentsline{toc}{chapter}{Introductory Remarks}
	\chaptermark{Introduction}
	\setcounter{chapter}{0}
	\setcounter{section}{0}
	\setcounter{figure}{0}
	\renewcommand{\thefigure}{0.\arabic{figure}}
	\markboth{Introductory Remarks}{Introductory Remarks}
	
	\begin{figure}[h]	   
	       \centering
	    % DO NOT ADD A FILENAME EXTENSION TO THE GRAPHIC FILE
	    \includegraphics[width=3.0in]{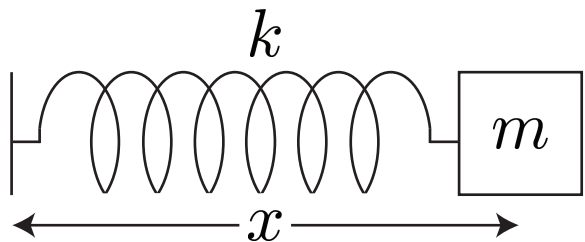}
	     \caption{The simple harmonic oscillator.}
	 \label{1spring}
	\end{figure}
Every freshman physics major is familiar with the simple harmonic oscillator.
First we introduce Hooke's law in one dimension, $F=-kx$, where $k$ is the {\em spring constant} and $x$ is the displacement from equilibrium. Then we invoke Newton's second law, $m \ddot x=\dot p=F$, where $\dot p$ is the total time derivative of momentum. From these we recover the equation of motion:
\begin{align*}
m \ddot x=-kx .
\end{align*}
We impose the initial conditions that the mass starts from rest, {\em i.e.} $\dot x(t)|_{t=0}=0$, and starts at initial displacement $a$, {\em i.e.} $x(t)|_{t=0}=a$. Using these we solve the ordinary differential equation and obtain the position as a function of time,
\begin{align}
x(t)= a  \cos \left(\sqrt{\frac{k}{m}} t\right) \label{shosolution}.
\end{align}
\begin{figure}[h!]	   
	       \centering
	    % DO NOT ADD A FILENAME EXTENSION TO THE GRAPHIC FILE
	    \includegraphics[width=6.0in]{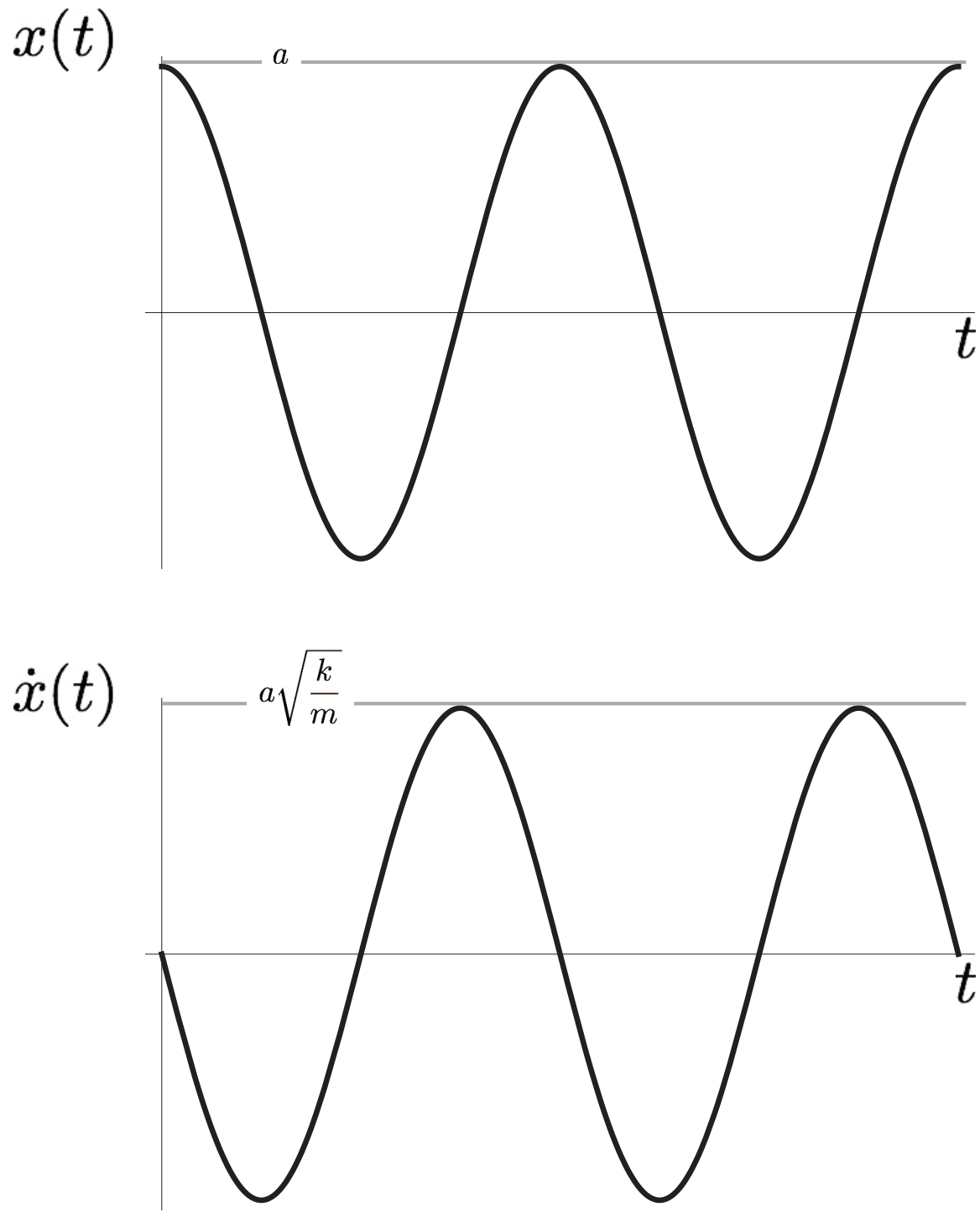}
	     \caption{Solutions to the simple harmonic oscillator.}
	 \label{sinecos}
	\end{figure}
It is here that we can pause to appreciate two qualities of interest: energy conservation and period which is independent of initial conditions.\newpage
Recall that potential energy $V$ of the system is $V= \frac{1}{2} k x^2$ and kinetic energy $K$ is \\$K=\frac{1}{2} m \dot x^2$, so the total energy of the system is
\begin{align*}
&\hspace{.85in}E=K + V = \frac{1}{2} k x^2+ \frac{1}{2} m \dot x^2.\\
\intertext{Plugging in our solution for $x(t)$ we can show that energy is conserved.}
E&=\frac{1}{2} k \left(a  \cos \left(\sqrt{\frac{k}{m}} t\right) \right)^2 + \frac{1}{2} m \left( a \sqrt{\frac{k}{m}} \sin \left( \sqrt{\frac{k}{m}} t \right)\right)^2=\frac{1}{2}k a^2,
\end{align*}
which is a constant, so $\frac{d E}{dt}=0$. As there is no change in energy with time, energy is conserved. 

Additionally, we find that the period is independent of initial conditions. By inspecting (\ref{shosolution}), we note that the simple harmonic oscillator has period given by 
\begin{align}
T=2\pi \sqrt{\frac{m}{k}}. \label{introshoperiod} 
\end{align}
This is only dependent on the parameters of the setup, $k$ and $m$, and is not dependent on the initial displacement $a$.

With the simple harmonic oscillator described, we now raise the concerns brought up by Einstein's special relativity. Recall that special relativity introduces two postulates, the principle of invariant light speed and the principle of relativity. While these postulates lead to a variety of interesting effects, we will focus on only a few features of special relativity which are relevant to our study. We should note that in Einstein's theory, energy $E$ is given by%%%
\begin{align*}
E=\sqrt{p^2 c^2 + m^2 c^4},
\end{align*}
where $p$ is relativistic momentum, and $c$ is the speed of light. We should also note that relativistic momentum is defined as
\begin{align}
p = \frac{m \dot x}{\sqrt{1-(\frac{\dot x}{c})^2}}\label{introrelp}.
\end{align}
For relativistic momentum we then have that as $\dot x \rightarrow \textrm{c}$, $p \rightarrow \infty$, such that exceeding the speed of light becomes impossible. So we get a universal ``speed limit,'' $c$. We also get a new, relativistic form of Newton's second law. From $\dot p = F$ we get
\begin{align}
\frac{d}{dt}\left \{ \frac{m \dot x}{\sqrt{1-(\frac{\dot x}{c})^2}} \right \} = F .
\end{align}
It is here that we begin to see conflicts with our current notion of the simple harmonic oscillator. Our solution (\ref{shosolution}) does not prohibit an oscillator which moves at a velocity greater than c. The use of relativistic momentum (\ref{introrelp}) will take care of this problem. 
 
 We should note that nothing---not even information---can travel faster than the speed of light. The linear restoring force we used earlier, $F=-kx$, assumes the instantaneous transfer of information along the spring. We will want to correct this by replacing this force with the gradient of a retarded potential, similar to what is done in electrodynamics.[1] 
 
 Additionally, special relativity introduces ambiguity in our linear restoring force: it is not clear in what reference frame $x$ is determined. Accordingly, we will explore a correction to the spring constant $k$ which arises when calculating the force due to the displacement in the rest frame of a mass and then transforming into an inertial ``lab frame.''

Let us proceed to reconcile these issues.
%%%%%%%%%%%%%%%%%%%%%%%%%%%%%% CHAPTER 1 %%%%%%%%%%%%%%%%%%%%%%%%%%%
%%%%%%%%%%%%%%%%%%%%%%%%%%%%%% CHAPTER 1 %%%%%%%%%%%%%%%%%%%%%%%%%%%
%%%%%%%%%%%%%%%%%%%%%%%%%%%%%% CHAPTER 1 %%%%%%%%%%%%%%%%%%%%%%%%%%%
\chapter{Relativistic Momentum}
\renewcommand{\thefigure}{\arabic{chapter}.\arabic{figure}}
We begin our examination of the relativistic spring by introducing relativistic momentum into Newton's second law. Doing so will ensure that the oscillator does not move at a velocity greater than $c$, our first grievance with the simple harmonic oscillator.

We will first show that this model conserves energy, then derive a nondimensionalized differential equation and proceed to discuss its numerically determined solutions. We will show that this correction gives a period dependent on initial conditions, unlike the non-relativistic result.
\vspace{-.016in}\section{The Model}
We invoke Hooke's law:
\begin{align}
\ttd{p}  &= - k x \text{ ,}\notag\\
\intertext{where $k$ is the spring constant, $x$ is the displacement from the origin, and $p$ is relativistic momentum. Recall that we have relativistic momentum}
p =& \frac{m \dot x}{\amma} \text{ ,}\notag\\
\intertext{where $m$ is the mass of the spring, and $c$ is the speed of light. We take then}
\ttd{}& \left \{  \frac{m \dot x}{\amma} \right \}  = -k x \text{ ,} \label{model1}
\end{align}
as our first statement of the model. We will proceed to discuss energy conservation and periodicity, and recover from (\ref{model1}) a second-order ODE which we will solve.

\subsection{Energy Conservation and Periodicity}
In order to arrive at our goal of an energy conservation statement, we wish to write this equation as a total time derivative. We will do so by first multiplying both sides by $\dot x$:
\begin{align}
\dot x \ttd{} \left \{ \frac{m \dot x}{\amma} \right \} = - k x \dot x \text{ .} \label{bothsidesbyxdot}
\end{align}
While the right side falls trivially,
\begin{align*}
-k x \dot x \;\; \rightarrow \;\; - \ttd{} \left \{ \frac{1}{2} k x^2 \right \} \text{ ,}
\end{align*}
the left hand side requires some coercion. With some perseverance (algebra), we find that
\begin{align*}
\dot x \ttdd{\gfrac{m \dot x}} \;\; \rightarrow \;\; \ttdd{\gfrac{m c^2}} \text{.}
\end{align*}
Putting the two results together with (\ref{bothsidesbyxdot}) gives
\begin{align}
\ttdd{\frac{1}{2} k x^2 + \gfrac{mc^2}}&=0 \text{.} \notag\\
\intertext{Integration by time, where we introduce constant $E$, yields}
\frac{1}{2} k x^2 + \gfrac{mc^2} =E, \label{energyconserv}
\end{align}
the desired energy conservation statement. And so, just like the non-relativistic simple harmonic oscillator, the model conserves energy.

Unlike the non-relativistic oscillator however, this model will not have a period independent of initial conditions.
If we assume $T$ is independent of $a$, we can choose an initial displacement $a$ such that $T<\frac{a}{c}$; as relativistic momentum requires $\dot x < c$, it is impossible for the oscillator to maintain period $T$. Thus period must depend on the initial conditions. 

We can recover from (\ref{energyconserv}) a second order ODE to probe these qualitative observations, and we will proceed to do so.
\subsection{Getting Our ODE}
Taking the total time derivative of our energy conservation statement (\ref{energyconserv}) gives us a second-order ODE,
\begin{align}
kx\dot x+\frac{m \dot x \ddot x}{(1-(\frac{\dot x}{c})^2)^{3/2}}&=0 \text{.} \label{secondorderode}
\end{align}
This is not analytically tractable, but can easily be handled using numerical techniques. In preparation for numerically solving, we nondimensionalize equation (\ref{secondorderode}). Let $t=t_o \tau$ and $x =x_o \overline x$, where $\tau \text{ and } \overline x$ are dimensionless variables and $t_o \text{ and } x_o$ are constants which carry the dimensions time and length respectively. Substituting these variables into our second-order ODE gives
\begin{align}
\overline x+ \frac{d^2 \overline x}{d\tau ^2}\left( 1-\frac{d\overline x}{d\tau}^2 \right)^{-3/2}&=0 \text{,} \label{nondimenrelode}
\end{align}
where it has made sense to let 
\begin{align*}
t_o = \sqrt{\frac{m}{k}}\text{ } \; \; \; \text{ and } \; \; \; x_o=\sqrt{\frac{m}{k}} c.
\end{align*}
Here $t_o$ is merely the fundamental time scale of the problem. Note that $t_o=\omega ^{-1}$, where $\omega$ is the frequency of oscillation in the non-relativistic SHO problem. From $t_o$ we then get the fundamental length scale of the problem $x_o$. From here we can proceed to examine the numerical solutions of (\ref{nondimenrelode}) obtained by {\em Mathematica's} Runge-Kutta implementation {\em NDSolve}.
%%%%%%%%%%%%%%%%%%%%%%%%%%%%%%%%%%%%%%%%%%%%%%%%%%%%%%%%%%%%%%%
%%%%%%%%%%%%%%%%%%%%%%%%%%%%%%%%%%%%%%%%%%%%%%%%%%%%%%%%%%%%%%%
%%%%%%%%%%%%%%%%%%%%%%%%%%%%%%%%%%%%%%%%%%%%%%%%%%%%%%%%%%%%%%%
%%%%%%%%%%%%%%%%%%%%%%%%%%%%%%%%%%%%%%%%%%%%%%%%%%%%%%%%%%%%%%%
\section{Solution}
\begin{figure}
  \centering
  \includegraphics[width=5.5in]{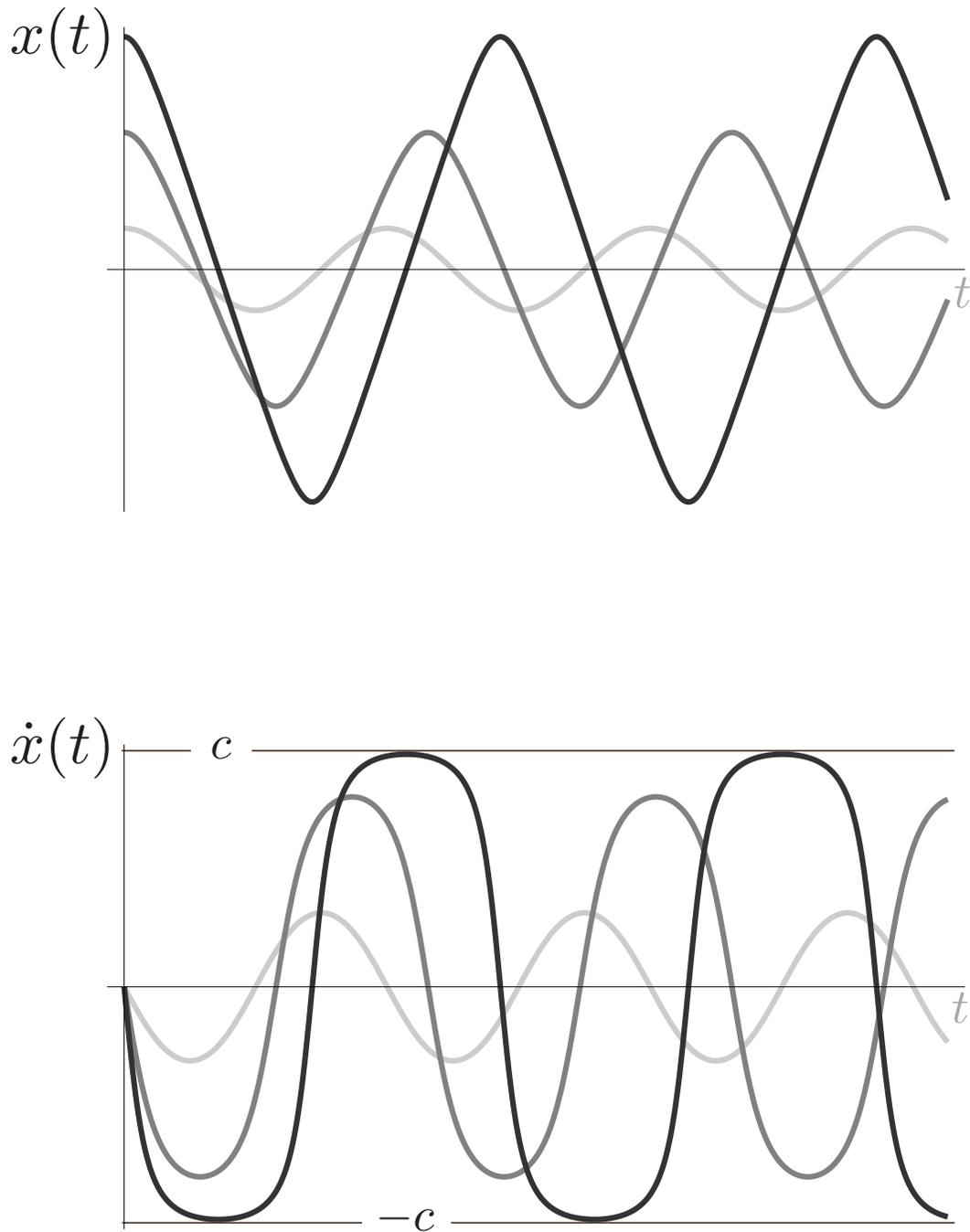}
  \caption[Results of numerically integrating the relativistic SHO.]% change eqn number...
  {Results of numerically integrating (\ref{nondimenrelode}). We note that the initial conditions vary from the classical limit (light gray), to the ultra-relativistic limit (black). Specifically: light gray: $a=.3 x_o$, gray: $a=1 x_o$, black: $a=1.7 x_o$.[2]  } \label{figrelode}
  \end{figure}
  Let's examine the solutions to (\ref{nondimenrelode}) with the initial conditions that the mass starts at rest with some initial displacement $a$ ({\em i.e.} $\frac{d\overline x}{d\tau}=0$, $\overline x(\tau=0)= a$). We will vary the initial displacement $a$ to observe the behavior of the oscillator as we move from the classical regime to the ultra-relativistic regime. We see the results of solving numerically by Runge-Kutta, courtesy of Wolfram's {\em Mathematica}, in figure \ref{figrelode}. The mass acts familiarly in the classical limit--when the initial displacement is significantly less than the fundamental length of the problem ({\em i.e.} $a \ll \sqrt{\frac{m}{k}}c$). However as we move into the relativistic regime, increasing the initial displacement to greater than the fundamental length $x_o$,  we see a shift in behavior. The velocity function begins to degenerate into a square wave as the mass spends more and more time at the maximum speed $c$. Accordingly, the position function begins to look like that of a photon bouncing between two mirrors, quickly changing direction when it reaches its maximum displacement, and heading back in the opposite direction at speed $c$. This aligns with our expectations, as the substitution of relativistic momentum explicitly imposes a speed limit on the model. That is, $p \rightarrow \infty$ as $\dot x \rightarrow c$. 

We also note that in the classical limit the period depends on $m$ and $k$, as expected. However, as the initial displacement increases, the period increases as well until it is entirely dependent on the initial displacement:
\begin{align}
T\;\; \rightarrow \;\;\frac{4a}{c}. \label{eqnrelperiod}
\end{align}
The period is simply the time it takes at speed $c$ to travel the full wavelength, that is, to go from $a$ to $-a$ and back to $a$. 

\begin{figure}
  \centering
  \includegraphics[width=5.5in]{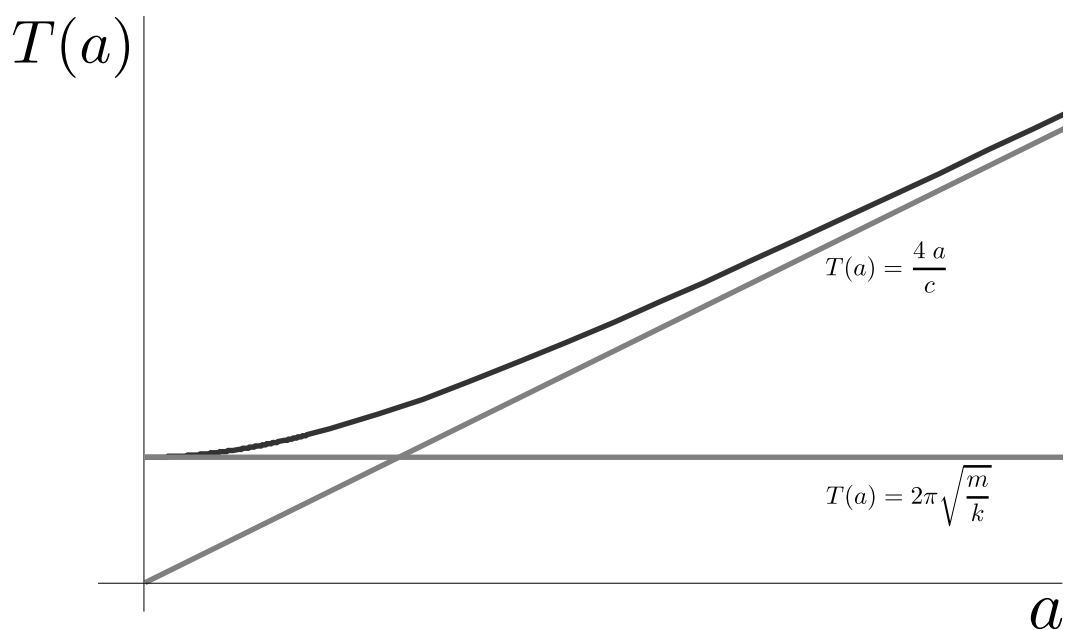}
  \caption[Period dependence of solutions to the relativistic SHO.]% change eqn number...
  {Plot of period $T$ by initial displacement $a$. We see can see the shift from the classical solution $T=2\pi \sqrt{\frac{m}{k}}$, which is independent of initial conditions, to the relativistic solution $T=\frac{4 a}{c}$.} \label{figrelperiod}
  \end{figure}

 We can watch this switch happen in our plot of period versus initial displacement in figure \ref{figrelperiod}. 
%%%%%%%%%%%%%%%%%%%%%%%%%%%%%% CHAPTER 2 %%%%%%%%%%%%%%%%%%%%%%%%%%%
%%%%%%%%%%%%%%%%%%%%%%%%%%%%%% CHAPTER 2 %%%%%%%%%%%%%%%%%%%%%%%%%%%
%%%%%%%%%%%%%%%%%%%%%%%%%%%%%% CHAPTER 2 %%%%%%%%%%%%%%%%%%%%%%%%%%%
\chapter{Delayed Forcing}
One of the most central ideas of special relativity is that nothing, not even information, can travel faster than the speed of light. The force law used in the last chapter, $F=\nobreak-kx$, assumes that the position information travels along the spring instantaneously. We must modify it and delay the force law. We will do so by replacing the force with the gradient of a retarded potential, similar to what is done in electromagnetism.
%%%%%%%%%%%%%%%%%%%%%%%%%%%%%%%%%%%%%%%%%%%%%%%%%%%%%%%%%%%%%%%
%%%%%%%%%%%%%%%%%%%%%%%%%%%%%%%%%%%%%%%%%%%%%%%%%%%%%%%%%%%%%%%
\section{The Model}
	\begin{figure}[h]	   
	       \centering
	    % DO NOT ADD A FILENAME EXTENSION TO THE GRAPHIC FILE
	    \includegraphics[width=4.0in]{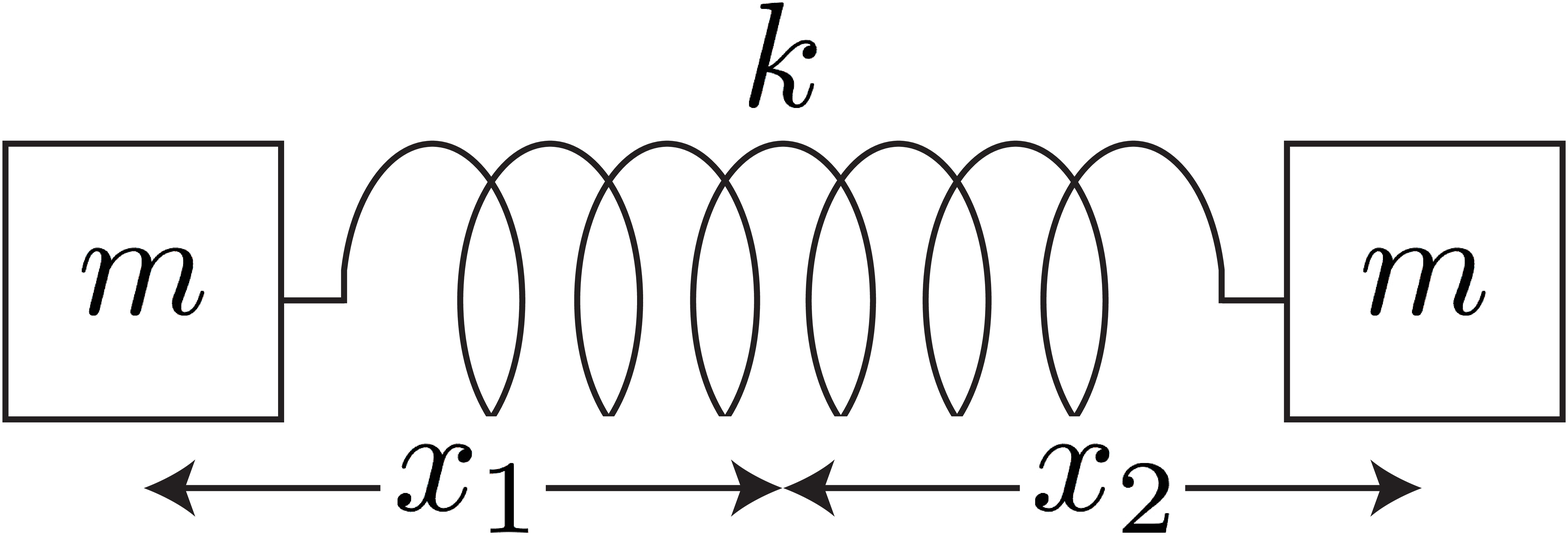}
	     \caption[Two masses connected by a spring.]{Two masses connected by a spring centered about the origin.}
	 \label{2spring}
	\end{figure}
Consider the model from the last chapter. We can think of the fixed point of the spring as the {\em source} of the spring potential and the mass as a {\em test particle} moving in the potential. Analogous to the force on one charge due to another in electrodynamics, the status of the ``source'' {\em right now} doesn't matter--for the spring potential ``news'' must travel at the speed of light. Instead, the status of the source some time in the past informs the potential. However, a fixed source particle (the wall in our prior model) subverts this problem by never changing its status, so to see the effects of this we must instead consider a different configuration: two masses connected by a spring. We will refer to them as the source mass and the test mass. The news takes a time $\frac{|x_{\textrm{t}}-x_{s}|}{c}$ to leave the source and reach the test particle, where $x_{\textrm{t}} \text{ and } x_{\textrm{s}} $ are the locations of the test mass and source mass respectively. We can then introduce the retarded time $t_{\textrm{r}}$ defined as
  \begin{align*}
t_r=t - \frac{| x_{\textrm{t}}(t) - x_{\textrm{s}}(t_r)|}{c}  \text{.}
\end{align*}
By this definition, if news arrives at $x_{\textrm{t}}(t)$, then $x_{\textrm{s}}(t_r)$ is the location of the source mass when the news left. Note that we cannot explicitly solve here for $t_r$ because this would require already knowing the functions $x_{\textrm{s}} (t)$ and $x_{\textrm{t}} (t)$.

Now we can simply plug $t_r$ into the Hooke's law potential to get the retarded potential,
\begin{align*}
V=\frac{1}{2} k \left( x_{\textrm{t}} (t) - x_{\textrm{s}} (t_r)\right)^2 .
\end{align*}
Taking the negative gradient of this potential should give us our force $F$. In this case, we simply take the derivative with respect to $x_{\textrm{t}}$:
\begin{align}
F=-k \left( x_{\textrm{t}} (t) - x_{\textrm{s}} (t_r)\right).
\end{align}
Using this force with Newton's second law, we create a set of coupled equations for the two masses. First we consider the assignment where, $m_1$ is the test particle and $m_2$ is the source; then we use $m_2$ as the test particle and $m_1$ as the source (figure \ref{2spring}). This gives us
\begin{align}
m \ddot x_1 &= -k (x_{1} (t) - x_{2} (t_r)) \label{delayodes1},\\
m \ddot x_2 &= -k (x_{2} (t) - x_{1} (t_r)). \label{delayodes2}
\end{align}
With symmetric initial conditions we only require one $t_r$ because the two masses are at every point indistinguishable, just reflected in space about the origin. Accordingly, to simplify our equations we will assume symmetric initial conditions and note that it is always possible to find a reference frame that has symmetric initial conditions. We choose retarded time
\begin{align}
t_r&=t - \frac{| x_{1}(t) - x_{2}(t_r)|}{c}.\label{delaytr}
\end{align}
Now we have a clear idea of our system, with the ODEs necessary for solving the trajectories of the masses. 

At the risk of sounding pedantic, we must make this clear: we are no longer describing the behavior of a true-to-life spring, in fact this system is a far cry from your everyday slinky; we are simply describing two point masses interacting via a delayed spring potential. Accordingly we must check our intuition at the door. We will proceed by discussing the qualitative features of this system.

\subsection{Energy Conservation and the Retarded Potential}
The system as described by (\ref{delayodes1}), (\ref{delayodes2}), and (\ref{delaytr}) appears to not conserve energy by construction. Consider a point of maximum displacement between the two masses, where we might naively assume we have maximum potential energy and no kinetic energy, as in the simple harmonic oscillator. However, in this formulation, at the maximum displacement we must have a finite retardation by definition, and so the potential is not yet at its maximum. Instead the potential will reach its maximum some finite amount of time after the maximum displacement, when the masses have accrued some amount of kinetic energy, and so the total energy at that time will be greater than the maximum potential energy. Because of this, even though we are deriving the force as the gradient of a potential, we cannot think of the force equation as being the time derivative of an energy conservation statement.

\subsection{Combination of ODEs}
To facilitate discussing the behavior of this system we combine our ODEs into one equation for the distance between the masses.  In order to do so we introduce displacement
\begin{align}
z(t)=x_1 (t) -x_2 (t).
\end{align}
Our equations (\ref{delayodes1}), (\ref{delayodes2}), and (\ref{delaytr}) become
\begin{align}
m \ddot z (t) &=-k(z(t) + z(t_r)) \text{,}\label{26}\\[4pt]
t_r&=t-\frac{1}{2}\frac{|z(t)+z(t_r)|}{c}. \label{dimendelay}
\end{align}

\subsection{Nondimensionalization}
Now we need to nondimensionalize these equations. Let $t=t_o \tau$, $x_1=x_o \overline x_1$, and $x_2=x_o \overline x_2$.
With these substitutions, our ODEs become
\begin{align*}
\frac{d^2 \overline x_1}{d \tau ^2} &= - (\overline x_1(\tau)+\overline x_2(\tau_r))\\[4pt]
\frac{d^2 \overline x_2}{d \tau ^2} &= - (\overline x_2(\tau)+\overline x_1(\tau_r)).
\intertext{Here}
\tau _{r} &= \tau - | \overline x_1(\tau) - \overline x_2(\tau_{r})|,
\end{align*}
and it has made sense to again assign
\begin{align*}
t_o = \sqrt{\frac{m}{k}}\text{ } \; \; \; \text{ and } \; \; \; x_o=\sqrt{\frac{m}{k}} c.
\end{align*}
Remember, $t_o$ is the fundamental time scale of the problem and $x_o$ is the length scale.
Similarly, for (\ref{dimendelay}) we let $t=t_o \tau$ and $z=x_o \overline z$, to get
\begin{align}
\frac{d ^2 \overline z}{d\tau^2} = - z(\tau)+z(\tau_r), \\[4pt]
\tau_r=\tau-\frac{1}{2}|z(\tau)+z(\tau_r)|.\label{nondimendelay}
\end{align}

\subsection{Choosing the Correct $\mathbf{\tau_r}$}
We note that with our definition in (\ref{nondimendelay}), we can have more than one value of $\tau_r$ for a given $\tau$.
The absolute value in our definition yields two possibilities:
\begin{align*}
\tau_r=\tau-\frac{1}{2}(z(\tau)+z(\tau_r)) \text{,} \; \; \; &\text{ and } \; \; \; \tau_r=\tau+\frac{1}{2}( z(\tau)+z(\tau_r))\text{.}
\intertext{From these we get}
\tau_r+\frac{1}{2}z(\tau_r)= \tau-\frac{1}{2}z(\tau)\text{,} \; \; \; &\text{ and } \; \; \; \tau_r-\frac{1}{2}z(\tau_r)=\tau+\frac{1}{2}z(\tau).
\end{align*}
Consider the functions $F(\tau)$, $G(\tau)$:
\begin{align*}
F(\tau)=\tau-\frac{1}{2}z(\tau) \text{,} \; \; \; &\text{ and } \; \; \; G(\tau)=\tau+\frac{1}{2}z(\tau).
\end{align*}
For general $z(\tau)$, $F$ and $G$ may be multivalued; we will show that this only occurs when the masses exceed speed $c$. For single valuedness we have the requirement that $F$ and $G$ are monotonically increasing functions (to avoid maxima and minima). For this we need $F'( \tau)\geq 0$ and $G'( \tau)\geq 0$, which is the case when $|\frac{dz(\tau)}{d\tau}|\leq 2$. This is true when neither mass exceeds the speed of light. Still, we have two possible values for $\tau_r$: one which is less than $\tau$ and one which is greater. We will choose strictly $\tau_r\leq \tau$ as the retarded time; the other option is the advanced time and is not physically relevant to us. We should note that when the masses exceed the speed of light---and they will until we re-impose the correct relativistic form of Newton's second law---we can have multiple values with $\tau_r\leq \tau$ and no clear idea of which to pick as the retarded time. We will see how this can incite some trouble when determining numerical solutions.
%%%%%%%%%%%%%%%%%%%%%%%%%%%%%%%%%%%%%%%%%%%%%%%%%%%%%%%%%%%%%%%
%%%%%%%%%%%%%%%%%%%%%%%%%%%%%%%%%%%%%%%%%%%%%%%%%%%%%%%%%%%%%%%
\section{Solution Method}
To deal with the functional dependence of our ODEs on $\tau_r$ we will implement {\em Verlet integration}. We will show that basic Verlet integration only requires a slight alteration to provide solutions using the retarded potential.[3]

\subsection{A Brief Aside on Verlet}
In the Verlet method we begin by discretizing time, taking a grid of size $\Delta \tau$. If we take the Taylor expansion of the position of one of our particles, $z(\tau)$, at two gridpoints $\tau_{+}=\tau+\Delta \tau$ and $\tau_{-}=\tau-\Delta \tau$, we obtain
\begin{align*}
z(\tau+\Delta \tau) &= z(\tau) + \dot z(\tau) \Delta \tau+ \frac{\ddot z(\tau) \Delta \tau^2}{2} + \frac{\dddot z (\tau) \Delta \tau ^3}{6} +\mathcal{O}(\Delta \tau ^4)\\
z(\tau-\Delta \tau) &= z(\tau) -\dot z(\tau) \Delta \tau+ \frac{\ddot z(\tau) \Delta \tau^2}{2}  - \frac{\dddot z (\tau) \Delta \tau ^3}{6} +\mathcal{O}(\Delta \tau ^4).
\intertext{Adding these together and solving, we can find an expression for $z(\tau+\Delta \tau)$}:
z(\tau+\Delta \tau) &= 2 z(\tau) -z(\tau-\Delta \tau) +\ddot z(\tau) \Delta \tau^2 +\mathcal{O}(\Delta \tau ^4).
\end{align*}
And so we get a function for $z(\tau+\Delta \tau)$ which is only dependent on $z(\tau)$, $z(\tau-\Delta \tau)$, and $\ddot z(\tau)$. With initial conditions and a description of $\ddot z$ we can take steps in time and plot the motion of our system.
%%%%%%%%%%%%%%%%%
\subsection{Computing $\mathbf{\tau_r}$}
Our function for $\ddot z(\tau)$ requires that we are able to find $\tau_r$, and now we will describe our method for doing so. By definition we have
\begin{align*}
\tau_r&=\tau-\frac{1}{2}|z(\tau)+z(\tau_r)|,
\intertext{which (as mentioned earlier) is not explicitly solvable for $\tau_r$. But we see that to discern the correct $\tau_r$, we can define the function $F$}
F(\sigma)&=\tau-\frac{1}{2}|z(\tau)+z(\sigma)| - \sigma ,
\end{align*}
which has the quality that $F(\sigma=\tau_r)=0$. In order to find $\tau_r$ we can repeatedly bisect the region $0\leq\sigma\leq\tau_r$, looking for the point where $F(\sigma)=0$, and taking $\sigma$ at the zero crossing of $F$ to be $\tau_r$. This assumes that there is only one $\tau_r \leq \tau$. In the case that the masses are moving faster than the speed of light we are {\em arbitrarily} choosing the bisection closest to current time $\tau$, which may lead to inaccuracies. However, we are not really interested in the solution in this case because we know it is unphysical.
%%%%%%%%%%%%%%%%%%%%%%%%%%%%%%%%%%%%%%%%%%%%%%%%%%%%%%%%%%%%%%%
%%%%%%%%%%%%%%%%%%%%%%%%%%%%%%%%%%%%%%%%%%%%%%%%%%%%%%%%%%%%%%%
\section{Solution}

%We should remark on some of the features of the solutions. Consider the mass starting from rest at some distance from the origin. After some finite has passed $|x(t)|<|x(t_r)|$, and the acceleration of a normal oscillator ({\em i.e.} a SHO, which does not exhibit growth) at that point is less than the acceleration due to the retarded force. So as the mass goes through it's first quarter period, it accelerates more than an SHO started at the same initial displacement, and thus has a greater velocity as it passes through zero displacement. At the zero-crossing--when there is no displacement--we note that $t_r=t$. As it moves away from zero, the finite displacement ensures that: $|x(t_r)|<|x(t)|$, and thus the mass is decelerated less as it reaches its new, greater maximum. This argument can be repeated so long as $t-t_r$ is no greater than the time it takes for a message to leave the last zero crossing, ({\em i.e.} so long as masses do not move with greater than c relative velocity). In the case that the masses do move greater than c, we begin to see far-lookbacks, which look further behind than the last zero crossing, and subsequently we observe chaotic growth and decay. 
\subsection{Growth and Periodicity}
\begin{figure}
  \centering
    \includegraphics[width=6.1in]{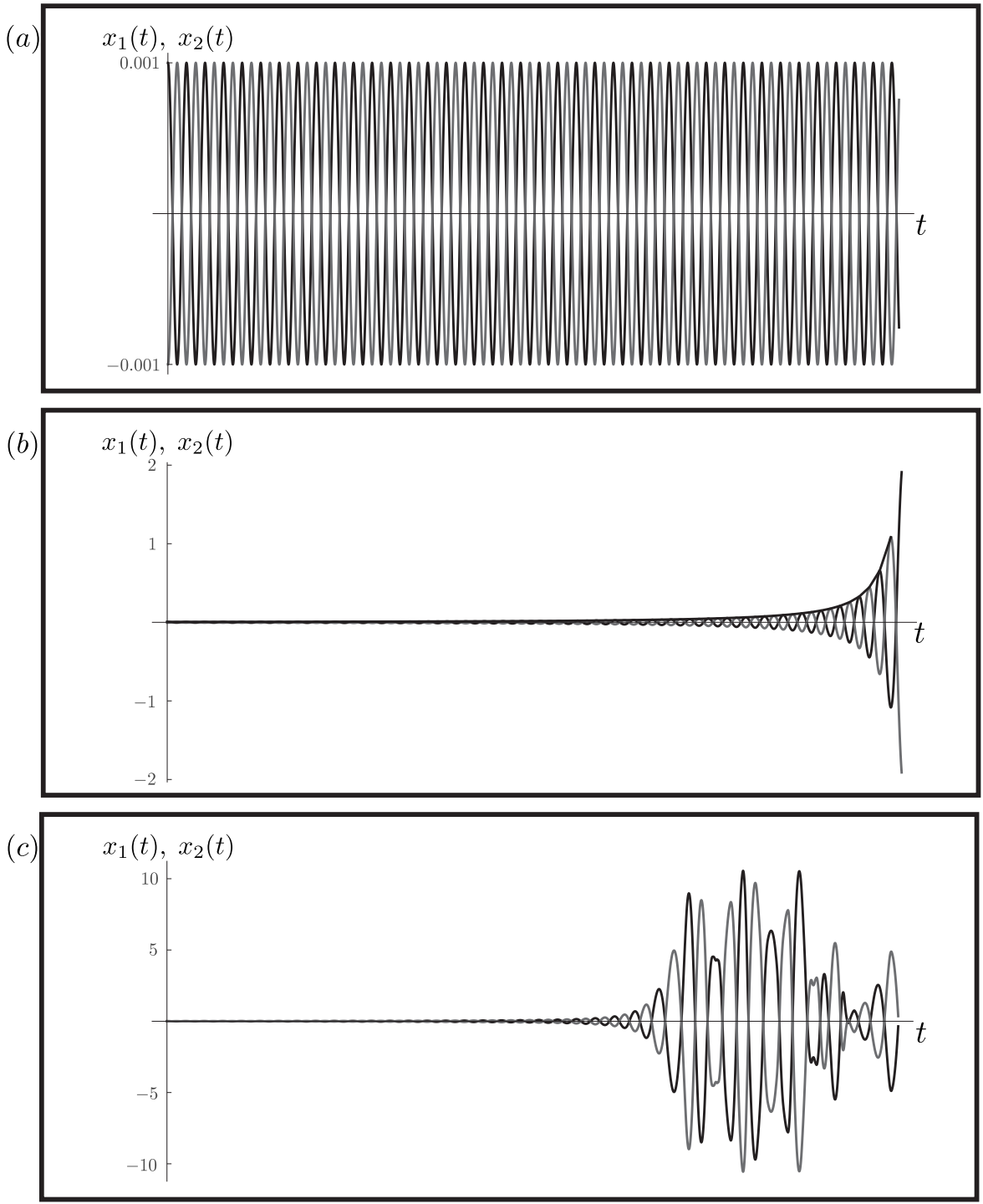}

  \caption[Verlet integration of the delayed force formulation.]{Results of our Verlet integration for ($a$) $a=0.001 x_o$, ($b$) $a=0.050 x_o$, and ($c$) $a=0.01 x_o$; $\Delta t = 0.005$; $35\;000$ steps. Trajectories of mass 1 and mass 2 (grey), the growth envelope of b (black line). Algorithm in appendix.} \label{trajectories}%The second figure depicts a starting position 0.5, 75000 steps at $\Delta t =$ 0.01. And the third figure depicts a starting position 0.05, 75000 steps at $\Delta t =$ 0.01  }
  \end{figure}

Figure \ref{trajectories} shows the results of our numerical methods for three different values of initial displacement. The solution with the smallest initial displacement is visually indistinguishable from the non-relativistic simple harmonic oscillator, but both of the others begin by exhibiting some growth. While we do not have an exhaustive explanation, we can at least qualitatively motivate the growth by comparing the retarded harmonic oscillator to the simple harmonic oscillator. Looking at some arbitrary point in the first quarter period of motion some finite time after releasing the mass from rest, we can see that the displacement ``seen'' by the retarded oscillator is greater than that seen by the simple oscillator. As this is true for the entire first quarter period, the retarded oscillator is accelerated more and is at a greater speed when it makes its first zero-crossing.

At the point of the first zero-crossing, $\frac{| x_{1}(t) - x_{2}(t_r)|}{c} \rightarrow 0$ and so $t_r\rightarrow t$. That is, the delay vanishes. Provided $\dot x < c$, after a zero-crossing the retarded oscillator cannot look back beyond the time at which the zero-crossing occurred $t_{\textrm{zero}}$. To see this we will focus our attention to some time after a zero crossing $t'=t_{\textrm{zero}} + \delta$. By definition we have the associated retarded time $t'_r$:
\begin{align}
t'_r=t_{\textrm{zero}} +\delta - \frac{| x_{1}(t') - x_{2}(t'_r)|}{c}.
\end{align}
We can preserve the ordering $t'_r \geq t_{\textrm{zero}}$ by requiring that $\delta \geq \frac{| x_{1}(t') - x_{2}(t'_r)|}{c}$. On the other hand we have a lower bound for the time elapsed since last zero-crossing: $\delta \geq \frac{x_1(t')}{\dot x_1(t_{\textrm{zero}})}$. Provided $\frac{x_1(t')}{\dot x_1(t_{\textrm{zero}})} \geq \frac{| x_{1}(t') - x_{2}(t'_r)|}{c}$, which is certainly the case if $\dot x < c$, we have $t'_r \geq t_{\textrm{zero}}$. This demonstrates that the retarded time for a given location never goes past the last zero crossing unless the mass moves faster than the speed of light.

At some arbitrary point in the second quarter period, we can see that the retarded oscillator ``sees'' a smaller displacement than the simple oscillator (provided it cannot look back past the last zero-crossing, {\em i.e.} $t'_r \geq t_{\textrm{zero}}$). As this is true for the entire second quarter period, the retarded oscillator is decelerated less and thus travels out to a greater maximum. For the beginning of the third quarter period, the retarded oscillator is looking back beyond the maximum and ``seeing'' a smaller displacement than the simple oscillator, and thus accelerating less. However, it is clear from the numerical solutions that this effect is not sufficient to prohibit growth. 

In figure \ref{trajectories} (c), we can see that once the oscillator begins to break the speed limit ($\dot x>c$), ``random lookbacks'' begin to occur. That is, the order $t'_r \geq t_{\textrm{zero}}$ is no longer preserved, and the retarded oscillator can look back arbitrarily far when determining the retarded force. This arbitrary force calculation makes for spontaneous growth and decay, which appears as an artifact of the oscillator moving faster than the information in the system. The seemingly random growth and decay is the result of demanding a speed limit but not enforcing it.

The important physical result of this section is the loss of energy conservation demonstrated by the growth of the oscillations. We can account for this by noting that we are not looking at a closed system. By imposing a retarded potential we have created the need for a field to mediate the spring force. As we are not accounting for the energy and momentum of this field, we have no sense of energy conservation in the system.

\begin{figure}
  \centering
    \includegraphics[width=6.1in]{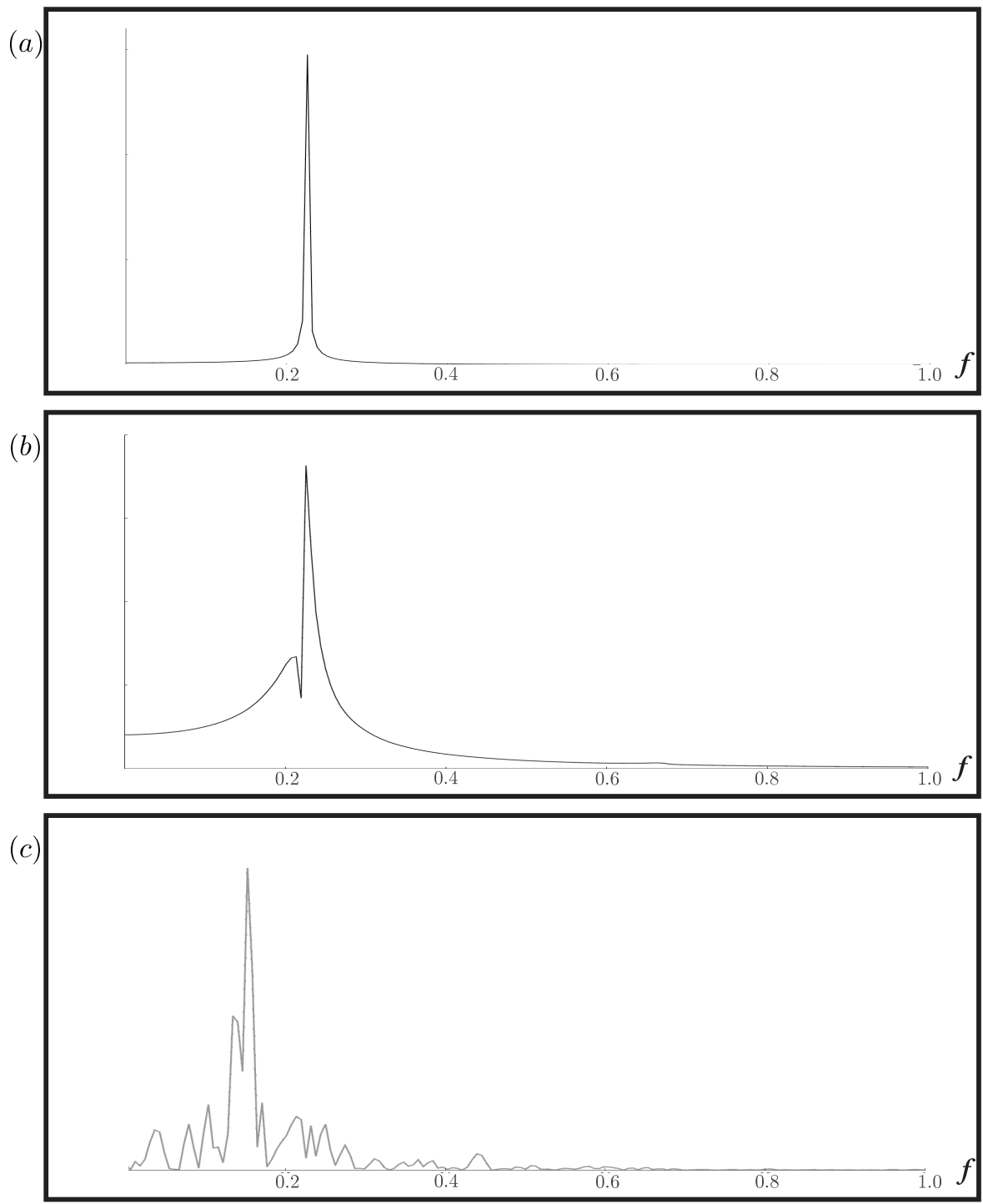}
  \caption[FFT of the Verlet integration solutions to the delayed force formulation]{Results of the FFT of our Verlet integration for ($a$) $a=0.001 x_o$, ($b$)\nobreak  $\;a=\nobreak0.050 x_o$, and ($c$) $a=0.01 x_o$; $\Delta t = 0.005$; $35\;000$ steps. Algorithm in appendix.}\label{retardedfft}
   %The second figure depicts a starting position 0.5, 75000 steps at $\Delta t =$ 0.01. And the third figure depicts a starting position 0.05, 75000 steps at $\Delta t =$ 0.01  }
  \end{figure}

Performing a fast Fourier transform on the numerical solution reveals that periodicity is independent of initial conditions and constant in time. Comparing the FFTs of the data in figure \ref{trajectories}, displayed in figure \ref{retardedfft}, we see that Fourier transforms have peaks at approximately $f=\sqrt{2}/2\pi$ Hz. This is precisely the frequency that we predict without any relativistic corrections (where the $\sqrt{2}$ comes from coupling the oscillators, see equation (\ref{26}) as $\tau_r\;\rightarrow\;\tau$). This is despite the vastly different initial conditions and behavior of the solutions ({\em i.e} the growth behaviors shown in figure \ref{trajectories}). In Chapter 1 we saw that relativistic momentum changes the periodic behavior without affecting energy conservation; now, we see that retarding the potential changes energy conservation without affecting the period. Naturally, our next endeavor will be combining relativistic momentum and the delayed force.

%%%%%%%%%%%%%%%%%%%%%%%%%%%%%% CHAPTER 3 %%%%%%%%%%%%%%%%%%%%%%%%%%%
%%%%%%%%%%%%%%%%%%%%%%%%%%%%%% CHAPTER 3 %%%%%%%%%%%%%%%%%%%%%%%%%%%
%%%%%%%%%%%%%%%%%%%%%%%%%%%%%% CHAPTER 3 %%%%%%%%%%%%%%%%%%%%%%%%%%%
\chapter{Putting It Together}
%%%%%%%%%%%%%%%%%%%%%%%%%%%%%%%%%%%%%%%%%%%%%%%%%%%%%%%%%%%%%%%
%%%%%%%%%%%%%%%%%%%%%%%%%%%%%%%%%%%%%%%%%%%%%%%%%%%%%%%%%%%%%%
In the past two chapters we introduced two corrections independently. First we introduced the relativistic correction to momentum, and then we formulated a delayed Hooke's law force. With the replacement of momentum by relativistic momentum, which put a restriction on the maximum speed allowed, we lost the period independence of initial conditions without affecting energy conservation. With our delayed force, which restricted the transfer of information in the system, we lost energy conservation without affecting period. Imposing both of these corrections simultaneously should produce interesting results. We predict that we will lose both energy conservation and simple periodicity.

 We will see the loss of energy conservation for the same reason as in the last chapter.  Even with the correction to momentum, we are not looking at a closed system, we are still neglecting to take into account the field which mediates the force. With the addition of the restriction of maximum speed $c$, we should again see periodicity which is not constant. In fact, coupling growth of amplitude with relativistic momentum, we should see growth of period over time.

\begin{figure}
  \centering
  \includegraphics[width=5.2in]{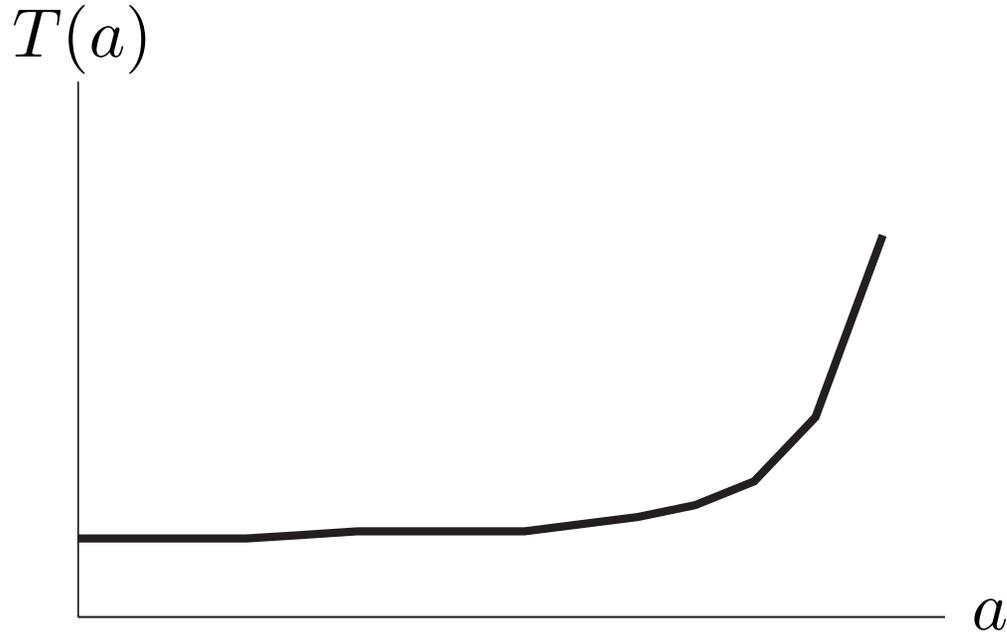}
  \caption[Prediction of period growth in the combination.]% change eqn number...
  {Our prediction from combining figure \ref{trajectories} and the function in figure \ref{figrelperiod}. See here the period shift that occurs over time due to growth; we should get longer and longer period.}\label{relperiodconvolution}
  \end{figure}

As we know the growth envelope of amplitude for a retarded potential from Chapter 2, and we know the period as a function of amplitude for the oscillator in Chapter 1, we should be able to predict the behavior of the amalgamation of the models. By plugging in the growth envelope of our results from figure\ref{trajectories} into the function in Fig \ref{figrelperiod} we can predict the changing period that should result in this final model. In \ref{relperiodconvolution} we show our prediction for period growth as a function of time.

In addition, it is interesting to include one final relativistic effect. One might take the position that the original simple harmonic potential $\frac{1}{2}kx^2$ should be interpreted as referring to the length $x$ in the rest frame of the test mass. In this case we get a transformation of the spring constant when moving to the inertial ``lab frame.''

\section{The Model}
Combining the delayed force and relativistic momentum, we get the system of equations
\begin{align}
\frac{dp_1}{dt}&=\frac{d}{dt}\left \{\frac{m \dot x_1}{\sqrt{1-(\frac{\dot x_1}{c})^2}} \right \} = -k(x_1(t)-x_2(t_r))\notag,\\
\intertext{ and} 
\frac{dp_2}{dt}&=\frac{d}{dt}\left \{\frac{m \dot x_2}{\sqrt{1-(\frac{\dot x_2}{c})^2}} \right \} = -k(x_2(t)-x_1(t_r)),\label{togethermodel}\\
\intertext{where}
t_r&=t-\frac{1}{2}\frac{|x_1(t)+x_2(t_r)|}{c}. \label{togethertr}
\end{align}

\subsection{Nondimensionalizing}
In order to solve (\ref{togethermodel}) and (\ref{togethertr}) we must nondimensionalize, as usual. Letting $t=t_o \tau$, $x_1=x_o \overline x_1$, $x_2=x_o \overline x_2$, $p_1=p_o \overline p_1$, and $p_2=p_o \overline p_2$, we have,
\begin{align}
\frac{d \overline p_1}{d \tau}&= -(\overline x_1(\tau)- \overline x_2(\tau_r))\notag\\[4pt]
\frac{d \overline p_2}{d \tau}&= -(\overline x_2(\tau)- \overline x_1(\tau_r)),\label{ch3nodimensn2l}
\intertext{where}
 \tau _{r} &= \tau - | \overline x_1(\tau) - \overline x_2(\tau_{r})|.\label{ch3nodimenstr}
\intertext{We define $p_1$ and $p_2$ as}
\overline p_1=\frac{\frac{d \overline x_1}{d\tau}}{\sqrt{1- \frac{d \overline x_1}{d\tau}^2}}&, \;\;\; \text{ and } \;\;\;\;
\overline p_2=\frac{\frac{d \overline x_2}{d\tau}}{\sqrt{1- \frac{d \overline x_2}{d\tau}^2}}.\label{ch3nodimensp}
\end{align}
In this case it makes sense to let
\begin{align*}
t_o = \sqrt{\frac{m}{k}}\text{, } \; \; \; \; x_o=\sqrt{\frac{m}{k}} c , \; \; \; \text{ and } \; \;\;  p_o=m c.\notag
\end{align*}
We see the time and length scale of the problem have stayed the same. Additionally, we introduce a natural momentum scale $p_o$ for the problem.
%%%%%%%%%%%%%%%%%%%%%%%%%%%%%%%%%%%%%%%%%%%%%%%%%%%%%%%%%%%%%%%
%%%%%%%%%%%%%%%%%%%%%%%%%%%%%%%%%%%%%%%%%%%%%%%%%%%%%%%%%%%%%%%
\section{Solution Method}
Because of the natural way in which equations (\ref{ch3nodimensn2l}) depend on momenta, it is valuable for us to develop a variation of the Verlet method in terms of momenta and positions. Here we will describe this.  

For clarity we will walk through the method with reference to one of the masses, $m_1$. Accordingly, we will make a few notational changes: we will refer to $x_1$ as $x$, $x_2$ as $y$, and $p_1$ as $p$. We will represent derivatives with respect to $\tau$ with the {\em dot} notation and we will throw out {\em barred} notation for nondimensional variables. All together, fore example, we will write $ \frac{d \overline x_1}{d \tau} \rightarrow \dot x$.

Our relevant equations are then, from (\ref{ch3nodimensn2l}) and (\ref{ch3nodimensp}),
\begin{align}
\dot p&= -( x(\tau)-  y(\tau_r))\\
p&=\frac{\dot x}{\sqrt{1- \dot x^2}}.\\
\intertext{We start with a grid in time with spacing $\Delta t$, as usual, where we define $x_j \equiv x(j \Delta t)$, $p_j \equiv p(j \Delta t)$, and $  t_r \equiv r\Delta t$.\newline
We note the definition of $p_j$,}
p_j&=\frac{\dot x_j}{\sqrt{1-\dot x_j^2}}.
\intertext{Inverting this for $\dot x_j$ and using the midpoint approximation of $\dot x _j$, we obtain}
 \frac{x_{j+1} - x_{j-1}}{2 \Delta t}&=\frac{p_j}{\sqrt{1+p_j^2}}.
\intertext{Solving for $x_{j+1}$ we get}
x_{j+1}&=x_{j-1}+\frac{ 2 \;p_j\Delta t}{\sqrt{1+p_j^2}}.
\intertext{This allows us to find $x_{j+1}$ given $x_{j-1}$ and $p_j$. To find $p_{j+1}$, we take the midpoint approximation of $\dot p_j$,}
\dot p_j &= \frac{p_{j+1}-p_{j-1}}{2 \Delta t},\label{311}
\intertext{where by the force equation, we have}
\dot p_j &= -( x_j-  y_r). \label{312}
\intertext{Together, (\ref{311}) and (\ref{312}) yield:}
p_{j+1} &= p_{j-1} -2 \Delta t \;( x_j-  y_r) .
\end{align}
So with formulae for $p_{j+1}$ and $x_{j+1}$ we can step in time to find our solutions, provided we can find $r\equiv t_r/\Delta t$. We will find $r$ by again calling a function $F$,
\begin{align*}
F(r)&= r \Delta t - j \Delta t + | \overline x_j - \overline x_r|,
\end{align*}
and running a bisection routine on $F$ to find $F(r \leq j)= 0$.

\section{Solution}
\begin{figure}
  \centering
  \includegraphics[width=5in]{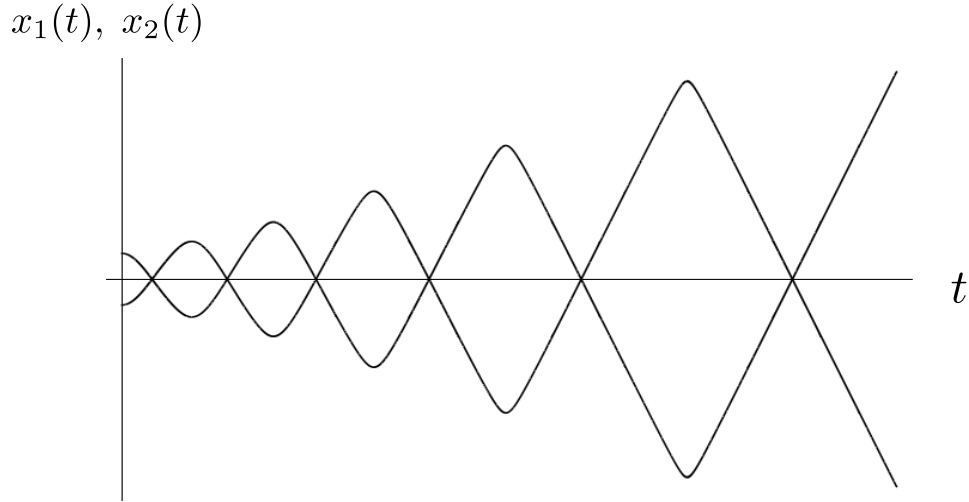}
  \caption[Qualitative example of combining effects.]% change eqn number...
  {Qualitative example of combining the effects, with $a=0.5 x_o$, $\Delta t = 0.001$, $30\;000$ steps.}\label{combo}
  \end{figure}
\begin{figure}
  \centering
  \includegraphics[width=6in]{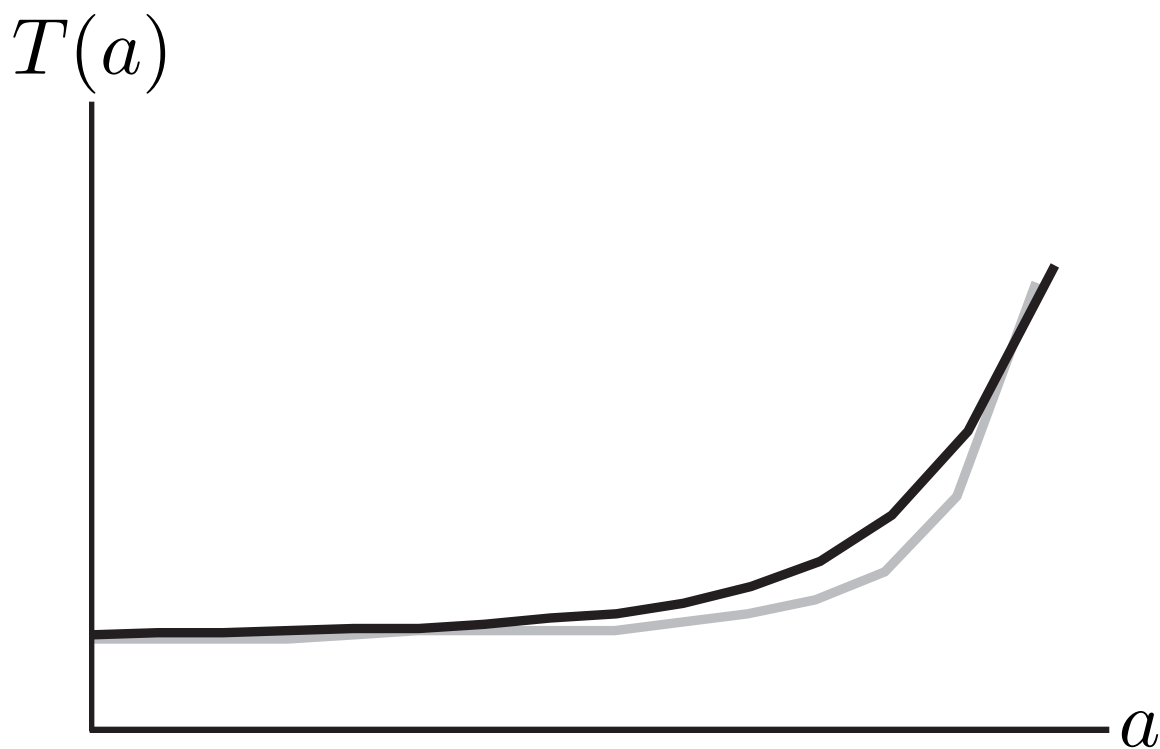}
  \caption[Convoluted prediction with actual period shift from numerics.]% change eqn number...
  {The prediction from figure \ref{relperiodconvolution} in grey, with period shift of solution in black.}\label{convolutionwithsolution}
  \end{figure}

The result of this numerical method is shown in figure \ref{combo}. We can immediately see the period growth and amplitude growth together, as was expected. As the oscillator grows in amplitude, it begins to travel at its maximum speed, and we see the shift from sinusoid to triangle wave. So the particles increase their maximum speeds until they look like a pair of photons bouncing between two mirrors as the mirrors move apart.
In figure \ref{convolutionwithsolution} we compare the period growth to that which we predicted based on a naive combination of the two corrections. While qualitatively the results are similar, they do have quantitative differences. \pagebreak
\section{Rethinking the Spring Constant}
\subsection{A Modification to our Model}
Consider for a moment how our equations must change as we change our reference frame. When we are in the test mass's rest frame we can calculate the total energy $\overline E$,
\begin{align*}
\overline E = mc^2 +\frac{1}{2} \overline{k x} ^2.
\end{align*}
Here, $\overline x$ is the separation of the two masses in the test mass's reference frame and $\overline k$ is the spring constant in this frame.
When we change to the lab frame (or any inertial reference frame), we have $E=\frac{\overline E}{\amma}$. To recover our formula for total energy in the \pagebreak[4] \\ lab frame, we must have
\begin{align*}
E&= \frac{ mc^2 +\frac{1}{2} \overline{k} \left(x \amma \right )^2 }{\amma }.
\intertext{This gives}
E&=\frac{mc^2}{\amma} + \frac{1}{2}k x^2,
\intertext{where }
k&=\overline k \amma .
\end{align*}\\
That is, the spring constant $k$ in the lab frame is different from the spring constant $\overline{k}$ in the test mass's frame. If we assume $\overline{k}$ is constant, this implies another modification to the force equation.
  With a corrected spring constant plugged into our solution, we can see that the growth in amplitude is slowed down (figures 3.4 and 3.5). However, the behavior is nearly identical.

\begin{figure}
  \centering
  \includegraphics[width=5in]{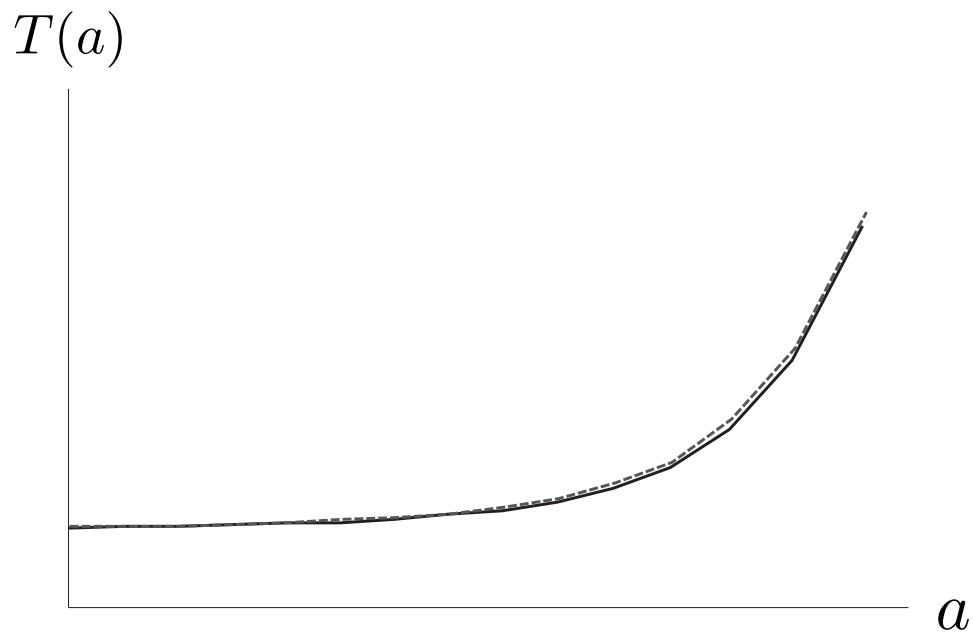}
  \caption[Period shift from numerics before and after spring constant correction.]% change eqn number...
  {Period shift from solution before (black) and after spring constant correction (black dashed).}
  \end{figure}
\begin{figure}
  \centering
  \includegraphics[width=5.5in]{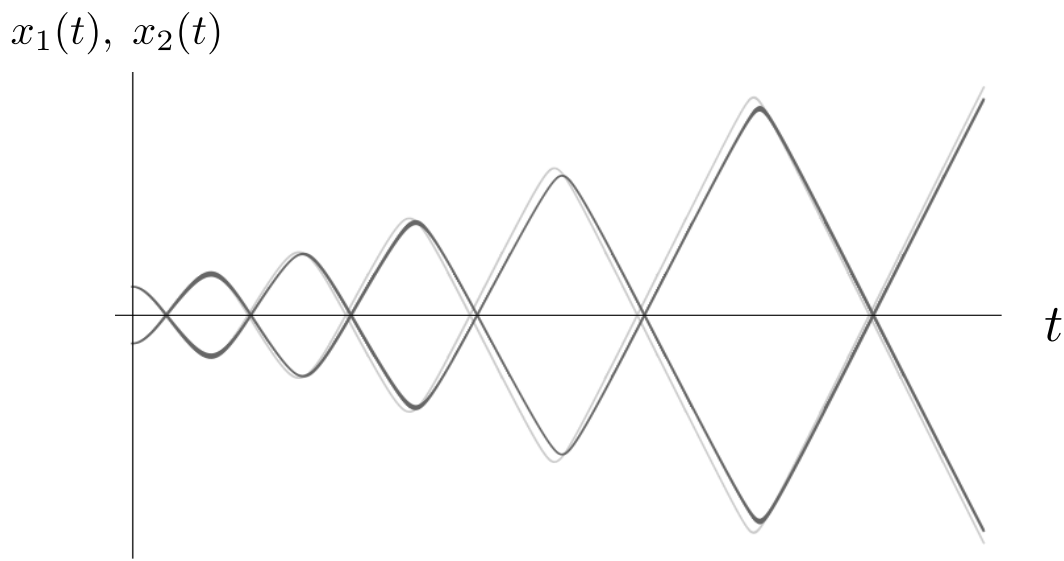}
  \caption[Qualitative example of spring constant correction.]% change eqn number...
  {Qualitative example of the correction applied to figure \ref{combo}, before correction (gray), after correction (black). Where, $a=0.5 x_o$, $\Delta t = 0.001$, $30\;000$ steps.}\label{combocorr}
  \end{figure}
 \chapter*{Closing Remarks}
 	\addcontentsline{toc}{chapter}{Closing Remarks}
	\chaptermark{Closing Remarks}
	\markboth{Closing Remarks}{Closing Remarks}
	\setcounter{chapter}{4}
	\setcounter{section}{0}
The simple harmonic oscillator undergoes a variety of relativistic corrections which alter its most familiar features such as energy conservation and a period independent of initial conditions. 

In Chapter 1 we implemented relativistic momentum to insure that the oscillator never moved faster than the speed of light. While this correction preserved energy conservation, it created a period dependent on amplitude. Imposing the ``speed limit'' while increasing the amplitude made the mass's trajectory look increasingly like that of a photon bouncing between two fixed mirrors. 

In Chapter 2 we introduced a delayed force in order to avoid having information move faster than the speed of light. While this correction had no affect on the period, it did cost us energy conservation. This delay implied that information propagate along the spring, but we weren't accounting for the energy of the field which mediated the force, and so we saw growth. 

In Chapter 3 we found that the amalgamation of these two effects gave us a prediction of how the combined model would act---an oscillator with increasing amplitude and period. Our prediction qualitatively lined up with what we observed numerically, although there were small differences. Finally we considered allowing the spring constant to transform with reference frame shift. This produced only slightly different numerical results. 

To continue this project it would be of value to continue stress testing the numerics of Chapter 3 and searching for the reason the prediction and numerics of chapter 3 do not perfectly align. As an extension, it may be advisable to formulate the field which governs the force described in Chapter 2. By doing so, and taking the dynamics of the field into account, it may be possible to recover energy conservation.

	\appendix
      \chapter{{\em Mathematica} Code}
    \section{Bisect While}
   This is the Code that is used to find the value of $\tau_r$:
\begin{verbatim}
BisectWhile[F_, maxz_, eps_] :=
  Module[{xm, xl, xr, vam, val, var, retval, i, pos, k},
   k = 0;
   (*If[maxz==1,Return[1]];*)
   retval = -1;
   xl = 1;
   xr = maxz;
   xm = Floor[(xl + xr)/2];
   val = F[xl];
   var = F[xr];
   vam = F[xm];
   If[Abs[F[xl]] > Abs[F[maxz]], pos = maxz + 1, 
    pos = 1];(*choose the smallest of the past or the current time*)
     If[val var > 0, retval = pos,(*if no zero crossings, 
    choose the smallest of the beginning or the current time*)
    While[Abs[vam] > eps && k < 10000,
     If[val vam < 0, xr = xm;, xl = xm;];
     xm = Floor[(xl + xr)/2];
     val = F[xl];
     var = F[xr];
     vam = F[xm];
     retval = xm;
     ];
    k = k + 1;
    ];
   (*first two fix the problem introduced by the floor function,
   	 i.e. look left and right. the third moves the retval to the
	    head if it has been flagged with "-1"*)
   If[retval == -1 | retval == xr, retval = retval, 
    If[Abs[F[retval + 1]] < Abs[F[retval]], retval = retval + 1]];
   If[retval == -1 | retval == xr, retval = retval, 
    If[Abs[F[retval - 1]] < Abs[F[retval]], retval = retval - 1]];
   If[retval == -1, retval = pos];
   Return[retval];
   ];
 \end{verbatim}
      \section{Verlet}
      This is the Code that is used to implement Verlet with the delayed force :
      	\begin{verbatim}
Verlet[xInitial_, deltaT_, N_] := 
 Module[{tol, i, tr, pos1, pos2, z, max, x1, x2, fVals},
  max = {{0, xInitial}};
  z = 0;
  tol = 10 deltaT;
  x1 = {xInitial, xInitial};
  x2 = {-xInitial, -xInitial};
  For[i = 2, i <= N, i = i + 1,
   fVals[j_] := 
    j deltaT - Length[x1] deltaT + Abs[x1[[Length[x1]]] - x2[[j]]];
   tr = BisectWhile[fVals, Length[x1] - 1, 10 deltaT];
   pos1 = (2 x1[[i]] - 
      x1[[i - 1]] - (deltaT^2) (x1[[i]] - x2[[tr]]));
   pos2 = (2 x2[[i]] - 
      x2[[i - 1]] - (deltaT^2) (x2[[i]] - x1[[tr]]));
   x1 = Join[x1, {pos1}];
   x2 = Join[x2, {pos2}];
   If[(x1[[i - 1]] - x1[[i - 2]])*(x1[[i]] - x1[[i - 1]]) < 0, 
    max = Join[max, {{deltaT*(i - 1), Abs[x1[[i - 1]]]}}]]];
  Return[{{x1, x2}, max}]]
	\end{verbatim}
	      \section{Momentum Verlet}
      This is the Code that is used to implement Verlet with the delayed force and Relativistic Momentum:
      	\begin{verbatim}
	ComboVerlet[xInitial_, deltaT_, N_] := 
  Module[{tol, i, pos1, pos2, z, zero, x1, x2, p1, p2, k, c, m, mom1, 
    mom2, j, fVals, tr},
   x1 = {xInitial, xInitial};
   x2 = {-xInitial, -xInitial};
   p1 = {0, 0};
   p2 = {0, 0};
   zero = {};
   For[i = 2, i <= N, i = i + 1,
    fVals[j_] := 
     j deltaT - Length[x1] deltaT + Abs[x1[[Length[x1]]] - x2[[j]]];
    tr = BisectWhile[fVals, Length[x1] - 1, 10 deltaT];
    mom1 = p1[[i - 1]] - 2 deltaT  (x1[[i]] - x2[[tr]]);
    mom2 = p2[[i - 1]] - 2 deltaT  (x2[[i]] - x1[[tr]]);
    pos1 = (2 p1[[i]]  deltaT)/Sqrt[1 + p1[[i]]^2] + x1[[i - 1]];
    pos2 = (2 p2[[i]]  deltaT)/Sqrt[1 + p2[[i]]^2] + x2[[i - 1]];
    x1 = Join[x1, {pos1}];
    x2 = Join[x2, {pos2}];
    p1 = Join[p1, {mom1}];
    p2 = Join[p2, {mom2}];
    If[x1[[i]]*x1[[i - 2]] < 0, zero = Join[zero, {deltaT*(i - 1)}]];
    ];
   Return[{{x1, x2}, zero}];
   ];
	\end{verbatim}
	
 \chapter*{Bibliography}
 	\addcontentsline{toc}{chapter}{Bibliography}
	\chaptermark{Bibliography}
	\markboth{Bibliography}{Bibliography}
	\setcounter{chapter}{5}
	\setcounter{section}{0}

\hangindent=.5in \hangafter=1
\newcommand{\bibfake}{\noindent \hangindent=.5in \hangafter=1}

\bibfake [1] D. J. Griffiths, \textit{Introduction to Electrodynamics} 3e (Prentice-Hall, New Jersey, 1999).\\

\bibfake [2] J. Franklin, \textit{Advanced Mechanics and General Relativity} 1e p.79 (Cambridge University Press, New York, 2010).\\

\bibfake [3] J. Franklin, course notes, \textit{Physics 367: Scientific Computation}, (Reed College, Spring 2011).
\end{document}